\documentstyle[12pt]{article}
\def\ns{\normalsize}

  \title{\Large\bf Signals of extra gauge bosons and exotic leptons in
                              SU(6)$_{L}\otimes$U(1)$_{Y}$ }

  \author{  Ricardo Gait\'an-Lozano$^{1,2}$, Albino
Hern\'andez-Galeana$^{1,4}$,\\
        Sergio A. Tom\'as$^{1}$, William A. Ponce$^{1,3}$ \\ and Arnulfo
Zepeda$^{1}$ \\[3em]
\ns 	1- Departamento de F\'\i sica, Centro de Investigaci\'on y de Estudios
Avanzados del IPN\\
\ns    A.P.~14--740, 07000 M\'exico D.F., M\'exico.\\
\ns     2- Departamento de F\'\i  sica, Universidad Surcolombiana,  A.~A.~385,
Neiva, Colombia.\\
\ns     3- Departamento de F\'\i sica, Universidad de Antioquia,  A.~A.~1226,
Medell\'\i n, Colombia.\\
\ns     4-Escuela de Ciencias, UAEM. Unidad del Cerrillo, Piedras Blancas\\
\ns         Instituto Literario 100, C.P 50000, Toluca Edo. M\'exico.}

     \newcommand{\bn}{\begin{eqnarray}}
     \newcommand{\en}{\end{eqnarray}}
     \newcommand{\be}{\begin{equation}}
     \newcommand{\ee}{\end{equation}}

\begin{document}
\thispagestyle{empty}
\renewcommand{\baselinestretch}{1}
\maketitle
\date{January 19, 1995}
\vskip.5cm

\begin{abstract}
{We study some of the consequences of the SU(6)$_{L}\otimes$U(1)$_{Y}$
model of unification of electroweak interactions and  families
with a horizontal gauge group SU(2)$_{H}$, paying special
attention to processes with flavor changing neutral currents.  We
compute at tree level the decays
$K^{+}\longrightarrow
\pi^{+}\mu^{+}e^{-}$, $K_L^0\longrightarrow \mu^{+}e^{-}$ and
$\mu^{-} \longrightarrow e^{-} \bar{\nu_e} \nu_\mu$ from which
we obtain lower bounds for the mass of the  horizontal gauge
boson associated with FCNC. Finally we obtain limits on the mixing between
ordinary and exotic
charged leptons. }
\end{abstract}
\pagebreak

\section{Introduction}
The precision measurements carried out in the last years have established
that the Standard Model (SM)$^{\cite{runo}}$ gives an excellent
description of the phenomena of particle physics up to 100 Gev. Even
though the SM describes with excellent accuracy the physics observed up
to now, it does not answer some questions such as the number of families,
CP violation and the fermion mass hierarchy problem. For these
reasons physicists believe that the SM is not the ultimate theory in
particle
physics. There are many attempts which try to give an answer to the above
questions named in a
generic form as extensions of the standard model. All of these extensions
imply the existence of new particles$^{\cite{rtres}}$ additional to those
introduced in the SM (either fermions, gauge bosons or scalars).
In some extensions the appearance of new exotic fermions is a necessary
condition to
free the model from anomalies, while the extra gauge bosons arise in
a natural way due to additional generators of the gauge group.
Exotic fermions manifest themselves either through direct
production or by their mixing with ordinary quarks and
leptons$^{\cite{rcuatro}}$.

In the SM there are several processes which are strongly
suppressed or forbidden; they are called {\it rare decays}. Of
special interest are those
due to the possible existence of
{\em flavor changing neutral currents} (FCNC). In
the SM the FCNC are suppressed by the Glashow-Iliopoulos-Maiani
(GIM) mechanism in the quark sector and by the conservation of
individual lepton numbers in the leptonic sector due to the masslessness of
the  neutrinos. The constraints on FCNC, and on rare decays in
general, play an
important roll in testing possible new physics beyond the SM.

Recently an extension of the SM, based on the
SU(6)$_L\otimes$U(1)$_Y$ has been introduced$^ {\cite{rcinco}}$ in
order to account for some of the peculiarities of the fermion
spectrum.  In particular the model is capable of accounting for
the fact that the top quark is much heavier than the rest of the
ordinary fermions.
In this article we study some of the new physics implied by this model;
in particular we analyze the combined effects due to the
appearance of new gauge bosons and exotic leptons.
We concentrate on FCNC processes which arise
in the model from three sources: horizontal interactions, mixing
between exotic and ordinary leptons, and mixing of the
standard neutral gauge boson with a horizontal one.

\section{The model}
The gauge group of the model is
SU(6)$_L\otimes$U(1)$_Y$, where SU(6)$_L$
unifies the weak isospin SU(2)$_L$ of the SM
with a horizontal gauge group G$_H$. SU(2)$_L\otimes$SU(3)$_H$ is a maximal
special subgroup of SU(6)$_{L}$.
The thirty-five SU(6)$_L$ generators in an SU(2)$_L\otimes$SU(3)$_H$
basis are
\be
\sigma_{i}\otimes {\bf 1}_{3},\hspace{2cm} {\bf 1}_{2} \otimes
\lambda_{\alpha},\hspace{2cm} \sigma_{i}\otimes \lambda_{\alpha},
\ee
where $\sigma_{i}$, i = 1,2,3 are  the Pauli matrices,
$T_{i}=\frac{1}{2}\sigma_{i}$ are the SU(2)$_L$ generators, $\lambda_{\alpha}$,
$\alpha=1,2,..,8$, are the Gell-Mann matrices, $\frac{1}{2}\lambda_{\alpha}$
are the SU(3)$_H$ generators, and
{\bf 1}$_{3}$ and {\bf 1}$_{2}$ are 3$\times$3 and 2$\times$2 unit matrices,
respectively.

The fermionic content of the model and the requirement that the exotics which
trigger the seesaw mechanism for neutrinos obtain mass at the scale where
SU(6)$_L$ is broken demand that G$_H$ be identified with the special maximal
subgroup SU(2)$_H$ of SU(3)$_H$ and not with SU(3)$_H$ itself. The special
maximal embedding of
SU(2)$_H$ into SU(3)$_H$ is achieved by using as generators of SU(2)$_H$ the
set
\be
(\lambda_{1} + \lambda_{6})/\sqrt2,
(\lambda_{2} + \lambda_{7})/\sqrt2, (\lambda_{3} +\sqrt{3}\lambda_{8})/2.
\ee
The three families are in the adjoint representations of SU(2)$_H$.

The model has 36 gauge bosons: 35 associated with the generators
of SU(6)$_L$ and 1 associated to U(1)$_{Y}$. Besides the standard
gauge bosons W$_{3}$, W$^{\pm}$ and B, we have 32 extra gauge
bosons, which can be divided in four classes: 12 charged gauge
bosons which produce transitions among families ({\em Family
Changing Charge Currents}, FCCC); 4 charged gauge bosons which
do not produce transitions among families but their couplings
are family dependent ({\em Non-Universal Family Diagonal Charged
Currents}, NUFDCC); 12 neutral gauge bosons which
produce
 transitions among families ({\em Flavor Changing Neutral
Currents}, FCNC) and 4 neutral gauge bosons which couple non-universally
without
changing flavor ({\em Non-Universal Flavor Diagonal Neutral
Currents}, NUFDNC).  The gauge bosons of SU(6)$_L$ and the
associated generators are displayed in Table 1.
The interactions mediated by the gauge fields associated
with the generators of SU(2)$_L$ are universal, that is, family independent.

With the usual definition for the electric charge operator $(Q=T_z +Y/2),$
the fermionic content of the model is given by the following set of
irreducible representations (irreps) of SU(6)$_L\otimes$U(1)$_{Y}$ (one for
each color in the case of quarks)
\begin{tabbing}
\{6(1/3)\}$_L$ \ \ \=
 = (u,d,c,s,t,b)$_L\equiv\psi^{\alpha}(1/3)_L$,\\
\{$\overline{6}(-1)\}_L$\>  = (e$^{-},\nu_{e},\mu^{-},\nu_{\mu},\tau^{-},
\nu_{\tau})_L\equiv\psi_{\alpha}(-1)_L$,\\
\{1$_{I}(-4/3)\}_L$\> = q$^{c}_{I}(-4/3)_L$ \ \ \ \=
I=1,2,3, for u$^{c}$, c$^{c}$, and t$^{c}$ respectively,\\
\{1$_{I}(2/3)\}_L$\> = q$^{c}_{I}$(2/3)$_L$ \> I=1,2,3 for
d$^{c}$, s$^{c}$, and
b$^{c}$  respectively,\\
\{1$_{I}(2)\}_L$\> = $l^{c}_{I}(2)_L$\> I=1,2,3
for e$^{c}, \mu^{c}$ and $\tau^{c}$  respectively,\\
\{$\overline{15}(0)\}_L$\> = $\psi_{[\alpha\beta]}(0)_L$,
\end{tabbing}
where I is a generational index, $\alpha$ and $\beta$ are SU(6)
tensor indices and $u,d,\ldots$ refers to the up quark, down
quark, $\ldots$ fields.  The label L refers
to left handed Weyl spinors and the upper c symbol indicates a
charge-conjugated field. The number in parenthesis is the
hypercharge and the symbol [$\alpha\beta$] stands for
antisymmetric ordering,
$\psi_{[AB]}=(\psi_{AB}-\psi_{BA})/\sqrt2$.

The particle content of $\psi_{[\alpha\beta]}(0)_L$
is 3 exotic electrically charged leptons $E_{1L}^{+}$, $E_{2L}^{+}$ and
$E_{3L}^{+}$ with their respective antiparticles $E_{1L}^{-}$, $E_{2L}^{-}$ and
$E_{3L}^{-}$, and 9 neutral Weyl states N$_i$.
Explicitly we write
\bn
\psi^{L}_{[\alpha\beta]}(0)=
\left(\begin{array}{cccccc}
         0 &N_{1} &E_{1}^{-} &N_{4}      &E_{2}^{-} &N_{6}\\
         {}&0     &N_{5}     &E_{1}^{+}  &N_{7}     &E_{2}^{+}\\
         {}&{}    &0         &N_{2}      &E_{3}^{-} &N_{8}\\
         {}&{}    &{}        &0          &N_{9}     &E_{3}^{+}\\
         {}&{}    &{}        &{}         &0         &N_{3}\\
         {}&{}    &{}        &{}         &{}        &0
      \end{array} \right)_L .                 \label{exotic}
\en
With respect to
SU(2)$_L$, the quantum numbers of the above exotic leptons
are 3 triplets and 6 neutral singlets. The three triplets are
\be
\left( \begin{array}{c} E_1^{+}\\ (N_4+N_5)/\sqrt2 \\ E_1^{-}
       \end{array} \right)\: , \hspace{.8cm}
\left( \begin{array}{c} E_2^{+}\\ (N_6+N_7)/\sqrt2 \\ E_2^{-}
       \end{array} \right)\: , \hspace{.8cm}
\left( \begin{array}{c} E_3^{+}\\ (N_8+N_9)/\sqrt2 \\ E_3^{-}
       \end{array} \right)\: ,
\ee
and the six neutral singlets are
\be
N_1, N_2, N_3, (N_4-N_5)/\sqrt2, (N_6-N_7)/\sqrt2,(N_8-N_9)/\sqrt2.
\ee

Possible deviations from the SM predictions could be either due to
mixing of ordinary with exotic leptons or due to mixings between the
standard gauge bosons and extra ones. Both effects modify the couplings
of ordinary fermions to the standard gauge bosons$^{\cite{rseis}}$.

\section{Symmetry Breaking.}
The symmetry breaking is realized in three stages: In the first stage
\be
\mbox{SU(6)}_L\otimes \mbox{U(1)}_Y \longrightarrow
\mbox{SU(2)}_L\otimes \mbox{SU(2)}_H \otimes \mbox{U(1)}_Y
\ee
at the scale M$_1$ where the six SU(2)$_L$ singlets of exotic leptons get mass
of order M$_{1}$. This breaking is achieved with a Higgs scalar in the
irrep $\phi_{1}=\{105(0)\}$ of SU(6)$_L$.

The next stage of symmetry breaking is
\be
\mbox{SU(2)}_L \otimes \mbox{SU(2)}_H \otimes \mbox{U(1)}_Y \longrightarrow
\mbox{SU(2)}_L \otimes \mbox{U(1)}_Y \;
\ee
at the scale M$_{2}$ and it is implemented with a Higgs
$\phi_{2}=\{\overline{15}(0)\}$.
At this stage the horizontal symmetry is completely broken
and simultaneously the exotic leptons which transform as triplets of
SU(2)$_L$ get a mass of order M$_{2}$.
The {\em Vacuum Expectation Values} (VEV's) of $\phi_{1}$ and $\phi_{2}$
can be read from Ref.\cite{rcinco}.
\vspace{.3cm}
\noindent
\paragraph{Charged ordinary fermion mass terms.}
The final stage of the symmetry breaking chain,
\be
\mbox{SU(2)}_L \otimes \mbox{U(1)}_Y \longrightarrow \mbox{U(1)}_{EM},
\ee
is achieved by using a Higgs $\phi_{3} = {\bar{6}(1)}\equiv \phi_{3\alpha}(1)$
with VEV's in the neutral components
$\langle \phi_{3\alpha}(1) \rangle = v_\alpha$ for $\alpha =1,3,5,$ (which
play the
same role as the ordinary Higgs in the SM model). This guarantees that
the only unbroken generator is Q.

With $\langle\phi_{3}\rangle$ the following mass term for quarks may be
written
\be                        \label{baremas}
\begin{array}{l}
\sum_{I} \gamma_{I}q^{cT}_L(-4/3)_{I}C \langle\phi_{3\alpha}(1)\rangle
\psi^{\alpha}_L
(1/3)+h.c.   \, ,                                              \\
=(\gamma_{u}u^{c}+\gamma_{c}c^{c}+\gamma_{t}t^{c})^{T}_LC
(v_{1}u+v_{2}c+v_{3}t)_L + h.c. \, ,
  \end{array}
\ee
where $\gamma_{I}$ are Yukawa couplings of order 1. The Q=2/3 quark mass
matrix at tree level produced by eq.(\ref{baremas}) is
\begin{equation}
\left(
 \begin{array}{ccc}
   \gamma_{u}v_{1} & \gamma_{u}v_{2} & \gamma_{u}v_{3} \\
   \gamma_{c}v_{1} & \gamma_{c}v_{2} & \gamma_{c}v_{3} \\
   \gamma_{t}v_{1} & \gamma_{t}v_{2} & \gamma_{t}v_{3}
 \end{array}
\right) \;
\end{equation}

\noindent
whose only
eigenvalue different from zero, $m_{t}=\gamma v$,
$\gamma=\sqrt{\gamma_u^2+\gamma_c^2+\gamma_t^2}$,
$v=\surd \overline{v_{1}^{2}+v_{2}^{2}+v_{3}^{2}}$,
may be
recognized as the bare top quark mass. Since we assume that the Yukawa
couplings are of order 1, then m$_t\sim$ M$_W=gv/2$. There are no
Q~=~$-1/3$ quark mass term but
$\phi_{3}$ allows also a mass term for the leptonic sector. This
mass term
would generate a very large mass for the
$\tau$ lepton. To avoid it we may either implement a universal see-saw
mechanism through the exotic charged leptons, or we may introduce a Z$_5$
discrete symmetry that
distinguishes between quarks and leptons as is explicitly done in
Ref.\cite{rcinco}.

The Higgses  introduced in this section  achieved the following tasks:
\begin{enumerate}
\item The VEV's $\langle \phi_{1} \rangle \oplus \langle \phi_{2} \rangle
     \oplus \langle \phi_{3} \rangle$
     break the symmetry SU(6)$_L\otimes$ U(1)$_{Y}$ down to U(1)$_{EM}$.
\item They produce heavy masses to all the exotic leptons.
\item They give a mass of order 10$^{2}$ GeVs to the t quark.
     The rest of the known fermion masses are light, because
     they can be generated only as radiative corrections$^{\cite{rsiete}}$.
\end{enumerate}

\section{Mixing between exotic and ordinary fermions}
To describe the mixing between ordinary and exotic fermions we follow
the formalism given in Ref.\cite{rseis}. Since U(1)$_{EM}$
is unbroken, different eigenstates of weak interactions can mix only when they
have the same electric charge.

In general the mixing among $n$ ordinary fermions and $m$ exotic
fermions of a given charge can be described by a unitary matrix of order
$(n+m)\times(n+m).$ It is convenient to introduce the vectors
$\psi_L^o$ and $\psi_R^o$ which can be decomposed in ordinary
and exotic sectors
\be
\psi_L^{o}= \left( \begin{array}{c} \psi_{EL}^{o}\\ \psi_{OL}^{o}
                   \end{array} \right)\; ,\hspace{2cm}
\psi_R^{o}= \left( \begin{array}{c} \psi_{ER}^{o}\\ \psi_{OR}^{o}
                   \end{array} \right)\; ,                        \label{euno}
\ee
where $\psi_{OL}^{o}$ is a column vector formed by $n_L$ ordinary fields, while
$\psi_{EL}^{o}$ contains $m_L$ exotic fields, and similarly, $\psi_{OR}^{o}$
contains $n_{R}$ ordinary fields, while $\psi_{ER}^{o}$ contains $m_{R}$ exotic
fields. The superindex $o$ refers to the weak interaction base, that is, to the
original fields in the lagrangean with well defined transformation properties
under SU(2)$_L\otimes$U(1)$_{Y}$. For each helicity one has $n+m$ mass
eigenstates, which can be arranged in vector columns as
\be
\psi_L= \left( \begin{array}{c} \psi_{hL} \\ \psi_{lL}
               \end{array} \right)\; , \hspace{2cm}
\psi_R= \left( \begin{array}{c} \psi_{hR} \\ \psi_{lR}
               \end{array} \right)\; ,
\ee
where $\psi_{lL}$ is a vector of $n_L$ light mass eigenstates, $\psi_{hL}$
represents $m_L$ heavy mass eigenstates, and similarly for $\psi_{lR}$,
$\psi_{hR}$.

The gauge and mass eigenstates are related by the equations
\be
\psi_L^o= U_L \psi_L \; , \hspace{2cm} \psi_R^o= U_R \psi_R,     \label{edos}
\ee
where $U_L$ and $U_{R}$ are unitary matrices of dimension $(n+m)\times(n+m)$,
which diagonalize the fermion mass matrices. $U_L$ and $U_{R}$ can be written
in  the block form
\be
U_a= \left(  \begin{array}{cc}    G_a & F_a \\
                                  E_a & A_a  \end{array}  \right)\, ,
                                  \hspace{1cm} a=L,R\, . \label{ua}
\ee

\noindent
Here $A_a$ is a mass matrix of dimension $n_a \times n_a$ which
relates the ordinary weak states and the light mass eigenstates
and $G_a$ is a matrix of dimension $m_a \times m_a$ which relates
the exotic and heavy states. $E_a$ and $F_a$ are of dimension
$n_a \times m_a$ and $m_a \times n_a$, respectively, and
describe the mixing of the two sectors. The matrices $A_a$ and
$G_a$ are not unitary, but from the unitarity of $U_a$ one obtains
the relations
\be
\begin{array}{lcc}
A_a^\dagger A_a +F_a^\dagger F_a &=& I_{n\times n} \, ,\\
A_a A_a^\dagger +E_a E_a^\dagger &=& I_{n\times n} \, ,\\
G_a^\dagger G_a +E_a^\dagger E_a &=& I_{m\times m} \, ,\\
G_a G_a^\dagger +F_a F_a^\dagger &=& I_{m\times m} \, .
\end{array}   \hspace{1cm} a=L,R.    \label{ru}
\ee
{}From these relations one can see that $A_a$ and $G_a$ violate
unitarity by a small mixing between light and heavy states.
Most of the physical effects of
mixing are related to the no unitarity of $A_a$.

An arbitrary mixing between exotic and ordinary fermions will
induce, in general, flavor changing neutral currents in the
light sector of fermions in a rate larger than the experimental
limits. These bounds show that for the charged sector these
processes of FCNC are extremely suppressed. Contrary to this situation,
for neutrinos there are no experimental limits on possible FCNC.
For this reason it is convenient to concentrate on the charged
fermions.

The present experimental limits$^{\cite{rdies}}$ on the transitions
$sd,cu, bd, bs, \mu e,\tau \mu$ and $\tau e$ allow us to assume
the non existence of FCNC involving
light fermions. The absence of FCNC combined with the fact
that the matrices $A^\dagger_aA_a$
and $F^\dagger_a F_a$ become diagonal, and that
$0\leq\left(A^\dagger_a A_a\right)_{ii}
\leq1$ and $0\leq\left(F^\dagger_aF_a\right)_{ii}\leq1$ allows one to
write$^{\cite{rcuatro}}$
\be
\begin{array}{lcl}
A^\dagger_aA_a &= &C^2_a \, , \\[.5em]
F^\dagger_aF_a &= &S^2_a \, ,
\end{array}
\qquad a=L,R.                             \label{edoce}
\ee
where
\be
\begin{array}{lcl}
C_a &= &{\rm diag}\;\left(C^1_a,C^2_a,\ldots,C^{n_a}_a\right),\\[.5em]
S_a &= &{\rm diag}\;\left(S^1_a,S^2_a,\ldots,S^{n_a}_a\right),
\end{array}
\qquad a=L,R
\ee
such that $C^i_a\equiv\cos\theta^i_a$ and $S^i_a\equiv\sin\theta^i_a$,
where $\theta^i_L$ and $\theta^i_R$ are the light-heavy mixing angles.
\vskip.3cm

\def\divis#1#2{{\displaystyle \frac{#1}{#2}}}
\section{Mass matrices}
In the model under study
the known charged fermions obtain mass only from radiative
corrections except for the top quark, which  obtains a mass at
tree level. The mass matrices of fermions and gauge bosons
obtained in the model have a complicated structure, making it
difficult to find an analytic solution to the eigenvalue problems.
To solve it we use an approximation method based on
 perturbation theory taking advantage of the fact that
the components of the mass matrices
depend on the three scales of SSB,  $M_{1}$, $M_{2}$ and $v$, with
the hierarchy $M_{1}\gg M_2 \gg v$.

\subsection{Mass for charged leptons}
{}From reference \cite{rcinco} we can write the following mass matrix for the
charged leptons in the base
$(E_1^0, E_2^0, E_3^0, e^0, \mu^0, \tau^0)$
\be
M^{(0)}_C= \pmatrix{M_2 &0 &0 &w &w &w\cr
                    0 &M_2 &0 &w &w &w\cr
                    0 &0 &M_2 &w &w &w\cr
                    0 &0 &0 &0 &0 &0\cr
                    0 &0 &0 &0 &0 &0\cr
                    0 &0 &0 &0 &0 &0\cr} \, .
\ee
where $w\ll v$, because this contribution comes from triplets of SU(2)$_L$.
The rank of this matrix is three and as a consequence it does
not give any mass to the electron, muon or tau. Adding the contribution
from radiative
corrections, $M_C$ can be parameterized as
\be         \label{radiativos}
M_C = \pmatrix{M_2 &0 &0 &w &w &w\cr
                    0 &M_2 &0 &w &w &w\cr
                    0 &0 &M_2 &w &w &w\cr
                    0 &0 &0 &V_1 &V_2 &V_3\cr
                    0 &0 &0 &V'_1 &V'_2 &V'_3\cr
                    0 &0 &0 &V''_1 &V''_2 &V''_3\cr} \; ,
\ee
where $V_i,V'_i,V''_i,i=1,2,3$, are real parameters of order of the electron,
muon and tau masses, respectively (at this stage we are ignoring phases,
so CP is conserved).

$M_C$ can be diagonalized by a biunitary transformation
\be
U_{CL}^\dagger M_C U_{CR}=M_D \; ,
\ee
where $M_D$ is a diagonal matrix and $U_{CL},U_{CR}$ are unitary matrices.
They satisfy
\be                            \label{diagon}
U_{CL}^\dagger M_C M^\dagger_C U_{CL}= U_{CR}^\dagger M^\dagger_C M_CU_{CR}=
 M_D^2 \, .
\ee

To diagonalize $M_C$ we write first
\be
M_C M^\dagger_C =\left( M_C M^\dagger_C \right)_0 +
                 \left( M_C M^\dagger_C \right)_1 \, ,
\ee
where $(M_C M^\dagger_C)_0=M_C^0M_C^{0\dagger}$, while
$( M_C M^\dagger_C)_1$ contains the parameters coming from radiative
corrections and their mixings with M$_2$ and $w$. $M_C^{(0)}$ is easily
diagonalized. Next we consider
$( M_C M^\dagger_C)_1$ as a perturbation. In order to
simplify the calculation we take $V_1^{''}=V_2^{''}=V_3^{''}\equiv V^{''}$.

The weak eigenstates  $e_L^0, \mu_L^0, \tau_L^0$
are related to the mass eigenstates $e_L, \mu_L, \tau_L$ through the
relations
\be
\begin{array}{rl}
e_L^0=& {\displaystyle {\sqrt3}} {\displaystyle \frac{w}{M^2_2}} (V_1 +V_2
+V_3)E_{1L} +\theta_1 e_L +\beta_1\mu_L +\alpha_1\tau_L,\\[1.3em]

\mu_L^0=& {\displaystyle {\sqrt3}}{\displaystyle \frac{w}{M^2_2}} (V'_1 +V'_2
+V'_3) E_{1L} +\theta_2 e_L +\beta_2\mu_L +\alpha_2\tau_L,\\[1.3em]

\tau_L^0= &{\displaystyle {\sqrt3}} {\displaystyle \frac{w}{M^2_2}}
V''E_{1L}+\theta_3 e_L +\beta_3\mu_L +\alpha_3\tau_L,
\end{array}
\ee
where the constants $\theta_{i},\beta_{i}$ and $\alpha_{i}$, i=1,2,3 are
functions of the parameters $V_{i},V_{i}^{'}$ and $V^{''}$.
The explicit expressions for $\theta_{i},\beta_{i}$ and
$\alpha_{i}$ are given in the Appendix.
The corresponding expressions for right handed charged leptons are not
needed here.

\subsection{Mass for neutral leptons}
With respect to the neutral sector, the masses for ordinary neutrinos are
generated through the see-saw mechanism. In the basis
\be
\begin{array}{c}
\left( N_1,N_2,N_3,\divis{(N_4-N_5)}{\sqrt2},\divis{(N_6-N_7)}{\sqrt2},
\divis{(N_8-N_9)}{\sqrt2}, \hspace{2cm} \right. \\
\left. \hspace{2cm} \divis{(N_4+N_5)}{\sqrt2},\divis{(N_6+N_7)}{\sqrt2},
\divis{(N_8+N_9)}{\sqrt2},\nu^0_e,\nu^0_\mu,\nu^0_\tau \right)\, ,
\end{array}
\ee
the mass matrix of this sector has the form
\be
M_{N\,\nu}= \left( \begin{array}{ccc}
                     A_{6\times 6}& 0_{6\times 3}& B_{6\times 3} \\
                     0_{3\times 6}& C_{3\times 3}& 0_{3\times 3} \\
                   B^T_{3\times 6}& 0_{3\times 3}& 0_{3\times 3}
                   \end{array}  \right)\, ,
\ee
where $A$ has entries of order $M_1$ and $M_2$, $B$ of order $v$,
and $C$ of order $M_2$. To diagonalize this matrix we use
a double perturbation theory$^{\cite{pzg}}$. After the algebra is done we
get the following expressions for the ordinary neutrinos
$\nu_e^0$ and $\nu_\mu^0$
\be
     \begin{array}{cl}
 \nu_e^0=&-\divis{1}{\sqrt2} \nu_1+\divis{1}{1.4668}\nu_2+
                            \divis{1}{5.3302}\nu_3+\sum
{\it O}(\delta_2)N_i \, ,\\
 \nu_\mu^0=&\divis{0.3842}{1.4668}\nu_2-\divis{5.1397}{5.3302}\nu_3
+\sum {\it O}(\delta_2)N_i \, ,
     \end{array}
\ee
where the parameter $\delta_2=v/M_2 \ll 1$. These expressions are used later to
compute the decay rate of the
muon.

\subsection{Neutral gauge bosons}
For the neutral gauge bosons (see Table 1) in the base
\[ \begin{array}{c}
\left( Z_0,Y_{H(1)},E_{13(1)},Y_{3(1)},E_{15(1)},\Sigma_2,Z_4,\Sigma_1,Z_3,
 Y_{r(1)},F_{13(1)} \right. \\
\left. Y_{3(2)},E_{15(2)},Y_{H(2)},E_{13(2)},Y_{r(2)},F_{13(2)} \right),
   \end{array}
\]
we obtain the mass matrix
\be \left(       \begin{array}{cc}
                     M^2_N  & 0    \\
                       0    & M^{'2}_N
                 \end{array}
    \right),
\ee
where $M^2_N$ is a symmetric mass matrix of dimension $11\times11$
containing terms of order $M^2_1$, $M^2_2$ and $v^2$; 0 are zero matrices
and $M'^2_N$ is a symmetric matrix of dimension $6\times6$.
$Z_0$ is the gauge boson of the SM,
\be
Z_0 ={\displaystyle \frac{gW_3-g'B}{\sqrt{g^2+g'^2}}},
\ee
where $g$ and $g'$ are the coupling constants of SU(6)$_L$ and U(1)$_Y$
respectively.

The diagonalization of $M^2_N$ is achieved considering
\be
    M^2_N = (M^2_N)_0 + (M^2_N)_1\, ,
\ee
with $(M^2_N)_0$ being the matrix with components of order $M_1^2$ and $M_2^2$
and $(M^2_N)_1$ containing components of order $v^2$. In a similar way to the
charged leptonic sector we solve the eigenvalues problem using
perturbation theory. The results show that $Z_0$ mixes  essentially with the
horizontal gauge boson $Y_{H(1)}$. This fact allow us to relate two mass
eigenstates, $Z_{SM}$ and $Z'$, with the eigenstates of weak interactions
through
an orthogonal transformation with a mixing angle $\Theta$, in the form
\be
\pmatrix{Y_{H(1)}\cr Z_0\cr} =
\pmatrix{\cos\Theta &-\sin\Theta\cr \sin\Theta &\cos\Theta\cr}
\pmatrix{Z'\cr Z_{\rm SM}\cr}
\ee
where
\be
\cos\Theta=\frac1{k_3},\quad \sin\Theta= \frac1{k_3}
\left(\frac2{3\sqrt3} \frac{v^2}{M^2_2} \frac{\sqrt{g^2+g'2}}g \right)
\ee
and
\be
k_3=\sqrt{1+ \left(\frac2{3\sqrt3} \frac{v^2}{M^2_2}
\frac{\sqrt{g^2+g'^2}}g \right)^2}
\ee
The mass of the $Z_{SM}$ gauge boson is given by
\be
M^2_Z \simeq\frac{M^2_W}{\cos^2\theta_W}
\ee
where $M_W$ is the mass of the charged boson W and $\theta_W$ is the weak
mixing angle given by
\be
\tan{\theta_W} = \frac{g'}{g} \, .
\ee
The mass for $Z'$ is of order $M_2$, while the mass for the other extra
gauge bosons are of order $M_1$ and $M_2$.
\vskip.5cm

\def\G{G_{\rm eff}}
\def\s{\section}
\def\r{\right}
\def\l{\left}
\def\no{\nonumber}

\def\sulh#1{SU(2)$_L\otimes$SU(#1)$_H$}
\def\su{{SU(2)$_L\otimes$U(1)$_Y$}}
\def\sc{{SU(3)$_C$}}
\def\sx{{SU(2)$_L$}}
\def\ssu{{SU(3)$_C\otimes$SU(2)$_L\otimes$U(1)$_Y$}}
\def\sg{{SU(2)$_L\otimes$U(1)$_Y\otimes$G$_H$}}
\def\ss{{SU(6)$_L\otimes$U(1)$_Y$}}
\def\st{\su}

\def\sus{SU(6)$_L\otimes$SU(6)$_C\otimes$SU(6)$_R\times$Z$_3$}

\def\mat#1#2{\pmatrix{#1\cr #2\cr}_L}

\def\bee{\begin{enumerate}}
\def\eee{\end{enumerate}}
\def\beq{\begin{eqnarray}}
\def\eeq{\end{eqnarray}}

\def\sus{SU(6)}
\def\suly{SU(6)$_L\otimes$U(1)$_Y$}
\def\ps{$Psi_{[AB]}$}
\def\tr{{\rm Tr}\;}
\def\uy{U(1)$_Y$}
\def\crd{{1\over\sqrt2}}
\def\sul{SU(6)$_L$}
\def\sulyd{SU(2)$_L\otimes$SU(1)$_Y$}
\def\sullh{SU(6)$_L\to$SU(2)$_L\otimes$SU(2)$_H$}
\def\suu{SU(3)$\otimes$SU(2)$\otimes$U(1)}
\def\ra#1{$\{#1\}$}

\def\e{\end}
\def\b{\begin}
\def\eq{{equation}}
\def\a{{array}}
\def\n{{eqnarray}}
\def\q{{eqnarray}}
\def\c{{center}}
\def\du#1#2{#1^{(#2)}_n}
\def\dd#1#2#3{#1^{(#2)}_{#3}}
\def\dt#1{{\cal L}^{(#1)}_{\rm lep}}
\def\dc#1#2{\big(N_{#1L}-N_{#2L}\big)}
\def\dcm#1#2{\big(N_{#1L} +N_{#2L}\big)}

\def\N#1{N_{#1L}}
\def\sqd{\frac{1}{\sqrt2}}
\def\ds#1#2{E^{#1T}_{#2L}CE^+_{#2L}}
\def\do#1#2#3#4{\sqd\dc{#1}{#2}^TC \sqd\dc{#3}{#4}}
\def\dom#1#2#3#4{\sqd\dcm{#1}{#2}^TC \sqd\dcm{#3}{#4}}
\def\divi#1#2{{\displaystyle \frac{#1}{#2}}}

\def\vn#1#2#3{ v'_{#1}N^T_{#2L}C\nu_{{#3}L}}
\def\vnd#1#2#3#4{ \sqrt2v'_{#1}\sqd \dc{#2}{#3}^TC\nu_{{#4}L} }
\def\xl#1#2{#1^{(0)}_{#2}}

\def\sq#1{{1\over\sqrt{#1}}}

\section{Calculation of decays}
In this section we compute the decays $K^+ \longrightarrow \pi^+\mu^-e^+$,
$K^0_L \longrightarrow \mu^+ e^{-}$
and $\mu \longrightarrow e \bar{\nu}_e \nu_{\mu}$.
The sensitivity of the experimental measurements allow us to give a lower
bound for the mass of the horizontal gauge boson $Z'$.

\subsection{$K^+ \longrightarrow \pi^+\mu^+ e^-$}

In this decay the only contribution to the amplitude arises from
the hadronic vector current, because the kaon and the pion have the same
parity.

The tree level diagram for  this decay is given by Fig.(1). The
decay amplitude is$^{\cite{nueveprima}}$
\be                                              \label{eseis}
\begin{array}{c}
{\cal M} = \divi{G_{\rm eff}}{\sqrt2} \bar e\gamma^\mu \big(C-D\gamma_5\big)
\mu A\l\{ f_+(t) \l[(P+P')_\mu -\divi{m^2_K-m_\pi^2}{t} (P-P')_\mu \r]
\r.\\[.5em]
\hspace*{\fill}\l.

+f_0(t)\divi{m^2_K-m^2_\pi}{t} (P-P')_\mu \r\} \, ,
\end{array}
\ee
where
\begin{flushleft}
\be  \label{econstants}
\begin{array}{lcl}
A & = &\divi14 \left[ V^*_{cs} V_{ud} +V^*_{ts} V_{cd} \right] (\cos\Theta+1)
 +\divi14 \left[ V^*_{us}V_{cd}+V^*_{cs} V_{td}\right](\cos\Theta -1)\\[.5em]
&&\mbox{}+ \sin\Theta \divi{C_{Vq}}{\sqrt3\cos\theta_W}  \left(
V^*_{us} V_{ud} +V^*_{cs}V_{cd} +V^*_{ts}V_{td}\right)\\[2em]
C & = & \sin\Theta {\displaystyle \frac{\delta_3}{\sqrt3 \cos{\theta_W}}} C_V
  + {\displaystyle \frac{\cos\Theta \Delta_1+\Delta_2}{4}} \, , \\  & &  \\
D & = & \sin\Theta {\displaystyle \frac{\delta_3}{\sqrt3 \cos{\theta_W}}} C_A
  + {\displaystyle \frac{\cos\Theta \Delta_1+\Delta_2}{4}} \, ,
\end{array}
\ee
\end{flushleft}
and
\be
\frac{G_{\rm eff}}{\sqrt2} = \frac{g^2}{8m_{Z'}^2}.
\ee
$V_{ab}$ are the elements of the Kobayashi-Maskawa mixing matrix and
\be    \label{definiciones}
\begin{array}{ccl}
\Delta_1& \equiv &
(\theta_1\beta_2+\theta_2\beta_1+\theta_2\beta_3+\theta_3\beta_2)\, ,\\[1em]
\Delta_2 & \equiv & (\theta_1\beta_2-\theta_2\beta_1+\theta_2\beta_3-
\theta_3\beta_2)\, ,\\[1em]
\delta_3&\equiv&(\theta_1\beta_1+\theta_2\beta_2+\theta_3\beta_3) \, ,
\end{array}
\ee
with $C_A$, $C_V$ and $C_{Vq}$ being the constants
\be     \label{cvyca}
    \begin{array}{ccl}
                 C_V & = & -1/2 + 2 \sin^2{\theta_W} \, ,\\
                 C_A & = & -1/2                      \, ,\\
              C_{Vq} & = & -1/2 + 2/3 \sin^2{\theta_W} \, .
    \end{array}
\ee
$P, P', P_1$ and $P_2$ are the fourmomenta of $K^+,\pi^+,\mu^+$ and $e^-$,
respectively, and $f_+(t)$ and $f_0(t)$ are form factors of order
1 (in fact, in our calculation they are taken equal to 1). Obviously
the momentum transfer $t=(P-P')^2$ is negligible compared with
the square mass of the horizontal gauge boson $Z'$, that is, $t\ll m_{Z'}^2$.

{}From the above amplitude we compute the partial decay rate
\be
d\Gamma =\frac1{(2\pi)^3} \frac1{32m^3_K} \sum_{\rm pol} |{\cal M}|^2 dt ds
\ee
with $s=(P-P_1)^2$, and we find
\be         \label{decayrate1}
\Gamma = \frac{9G^2_F(C^2+D^2)A^2}{32(2\pi)^3}
\frac{M^4_W}{m_{Z'}^4 m^3_K} \big[2.464\times 10^{20}\ {\rm MeV}^8\big]
\ee
where $G_{\rm eff}$ and $G_{F}$ (Fermi's constant) are related by
\be
G_{\rm eff}^2 = 9\frac{M^4_W}{m_{Z'}^4} G^2_F.
\ee

As we can see, $\Gamma$ depends on several parameters of the model. Among
these parameters we have $\theta_i, \beta_i$ and $\alpha_i$ (which are
functions of $V_i$, $V'_i$ and $V''_i$ introduced from the radiative
corrections generating the masses of charged leptons, eq.(\ref{radiativos})),
the angle $\Theta$ (which describes the mixing between the gauge bosons
$Z_0$ and $Y_{H(1)}$) and the mass $M_{Z'}$ of the horizontal boson $Z'$.

In order to compare the expression (\ref{decayrate1}) with the experimental
limits$^{\cite{rnueve}}$, we have taken
\be                              \label{aproximation}
      \begin{array}{c}    V_i  \sim m_e    \, ,\\
                          V'_i \sim m_\mu  \, ,\\
                          V''  \sim m_\tau \, .
      \end{array}
\ee
With the above approximations the only non negligible parameters in the set
($\theta_i, \beta_i$ , $\alpha_i$) are $\theta_1$, $\beta_2$ and
$\alpha_3$. In particular
\be
(\theta_1 \beta_2)^2 \sim (1 + \frac{m_\mu^2}{m_\tau^2})^{-1} \, .
\ee
In consequence we obtain the values
\be          \label{cyd}
\begin{array}{ccl}
C &= &-0.4927  \, ,\\[.5em]
D &= & 0.4927  \, ,\\[.5em]
A &= & 0.2300  \, .
\end{array}
\ee
Taking into account the values for $m_\mu$,
$m_K$,
$m_\pi$,
$M_W$,
$G_F$ and the experimental bound for this decay$^{\cite{rdies}}$
\[
\Gamma^{\rm exp} < 11.17327314\times10^{-24}\ \rm Mev,
\]
we get the constraint
\be
m_{Z'} > 19.1767 \; {\rm TeV.}
\ee
\subsection{$K^0_L\longrightarrow e^-{\mu^+}$}

The state $K^0_L$ is a superposition of the states $K^0$ and
$\bar K^0$, defined as
\be
|K^0_L\rangle =\frac{ \l[ (1+\epsilon)|K^0\rangle -(1-\epsilon)|
\bar K^0\rangle \r]}   {\l[2(1+|\epsilon|^2)\r]^{1/2} } \; ,
\ee
$\epsilon$ being a parameter that describes the CP violation
mixing.

The amplitude for the leading diagram contributing to the process,
Fig. (2), is given by

\be
{\cal M}=-\frac{\G}2 \bar e \gamma^\mu \big(C-D\gamma_5\big) \mu B f_K
\big(P_1+P_2\big)_\mu.         \label{veinticinco}
\ee
where C and D are given by eq.(\ref{econstants}) and
\be
\begin{array}{lcl}
B &= &-\divi14 \left[V^*_{cs}V_{ud}+V^*_{ts}V_{cd}\right] (\cos\Theta+1)
 -\frac14 \left[V^*_{us}V_{cd} +V^*_{cs} V_{td}\right] (\cos\Theta -1)\\[.5em]
  &  & + \sin\Theta \divi{1}{2\sqrt3\cos\theta_W}
      \left( V^*_{us} V_{ud} +V^*_{cs}V_{cd} +V^*_{ts}V_{td}\right) \,
\end{array}
\ee
$P_1$ and $P_2$ are the fourmomenta of the electron and the muon
respectively,
and $f_K$ is the decay constant of the kaon. In the above amplitude we have
considered again that the momentum transfer is small compared with the square
mass of the boson $Z'$. Hence we obtain the total decay rate
\be
\Gamma = \frac{9B^2f^2_{K}}{32\pi} \frac{M^4_W}{m_{Z'}^4} G^2_F
\big(C^2+D^2\big) \frac{m^2_\mu}{m^3_{K^0}} \big(m^2_{K^0}-
m^2_\mu\big)^2.
\ee
Now, taking into account the approximations given in Eq. (\ref{aproximation})
we find
\be
B = -0.2300 \,.
\ee
Using now the experimental data $^{\cite{rdies}}$
\be
\begin{array}{lcl}
f^+_K &= &160.6 \ \rm MeV                \, , \\[.5em]
m_{K^0} &= &497.676\pm0.030\ \rm MeV     \, , \\[.5em]
\Gamma^{\rm exp}& < & 12.01397025\times 10^{-25}\ \rm MeV  \, ,
\end{array}
\ee
and taking for the parameters $C$ and $D$ the values given in eq. (\ref{cyd}),
we obtain the constraint
\be  m_{Z'}  > 34.7546 \: {\rm TeV}  \ee


\subsection{Muon Decay $\mu \longrightarrow e \bar{\nu}_e \nu_\mu.$}

Another process that allow us to establish a lower bound for the mass of the
horizontal gauge boson $Z'$ is the muon decay
$\mu \longrightarrow e \bar{\nu}_e \nu_\mu.$ This process
is mediated by the standard W charged boson as well as by the
horizontal boson $Z'$. The corresponding diagrams are shown in Figs. (3) and
(4).

The total amplitude is
\be
{\cal M}={\cal M}_W + {\cal M}_{Z'}
\ee
where ${\cal M}_W$ (${\cal M}_{Z'}$) is the amplitude for  the process
mediated by the $W$ ($Z'$). In order to get the couplings we must
take into account that, unlike the charged sector, in the neutral
sector we do not know the masses for the neutrinos. This fact gives
as a result that, while the charged leptons are written in the lagrangean
in terms of their mass eigenstates, the neutrinos are left in the weak
interaction base.

The amplitudes ${\cal M}_{Z'}$ and ${\cal M}_W$ are
\beq  {\cal M}_W =  \left[ G_\mu \overline{\nu_\mu} \gamma\,^\mu
\left( {\displaystyle \frac{1-\gamma^5}{2}} \right) \mu \right]
{\displaystyle \frac{g_{\mu \nu}}{M_W^2}}
                     \left[ G_e   \overline{e} \gamma^\nu
\left( {\displaystyle \frac{1-\gamma^5}{2}} \right) \nu_e \right] \, , \eeq
and
\be
{\cal M}_{Z'}= -\left[ G_n \overline{\nu_\mu} \gamma\,^\mu
\left( {\displaystyle \frac{1-\gamma^5}{2}} \right) \nu_e \right]
{\displaystyle \frac{g_{\mu \nu}}{M_{Z'}^2}}
                      \left[ G_{ch} \overline{e} \gamma^\nu
\left( {\displaystyle \frac{C'_V-C'_A\gamma^5}{2}} \right) \mu \right] \, ,
\label{mzprima}
\ee
where
\be G_n \equiv {\displaystyle \frac{g}{2\sqrt2}}
{\displaystyle \frac{\cos\Theta+1}{2}} \, ,\hspace{2cm}
   G_{ch} \equiv {\displaystyle \frac{g}{2\sqrt2}}
{\displaystyle \frac{\cos\Theta \Delta_1+\Delta_2}{2}} \, ,
\ee
and
\be G_\mu \equiv
\frac{g_{_{SM}}}{\sqrt2}
\beta_2 \, ,\hspace{1cm}       G_e \equiv
\frac{g_{_{SM}}}{\sqrt2}
\theta_1 \, , \hspace{1cm} g_{_{SM}} = g/\sqrt{3}. \,  \ee
We have defined
\be   \begin{array}{ccl}
C'_V&\equiv& x C_V + 1 \, , \\  & &  \\
C'_A&\equiv& x C_A + 1 \, , \\ & & \\
x&\equiv& {\displaystyle \frac{4\sin\Theta}{\cos\Theta \Delta_1+\Delta_2}}
{\displaystyle \frac{\delta_3}{\sqrt3 \cos{\theta_W}}} ,
\end{array}
\ee
where we have used the relations given by (\ref{definiciones})
and (\ref{cvyca}) and we have considered that the transferred momentum  in
the propagator  is small compared with
$M_{Z'}^2\:(M_W^2)$.

To evaluate the
nonstandard contribution to the amplitude, given by Eq.\ref{mzprima},
we Fierz-transform it and compare its scalar and pseudoscalar
part with the parameterization used in Ref. \cite{mzprima}
\begin{equation}
{\cal M}_{Z'}^{SP} = \frac{g^{2}}{2\sqrt{2} M_W^2} g^{-+}
\bar{e}_{R} \nu_{eL} \bar{\nu}_{\mu L} \mu_{R}
\end{equation}
\noindent
and we obtain
\begin{equation}
\begin{array}{ccl}
g^{-+}& = & \frac{G_{n}G_{ch}}{g^2} 8\sqrt{2} x \frac{M_W^2}{M_{Z'}^2}
\sin^2 \Theta_W \\ & & \\
      & = & \sqrt{6} (\cos\Theta + 1)\sin\Theta \delta_3
\frac{\sin^2\theta_W}{\cos\theta_W} \frac{M_W^2}{M_{Z'}^2}.
\end{array}
\end{equation}
Using now the bounds for $g^{-+}$ obtained in Ref. \cite{fgj}, we get
\begin{equation}
M_{Z'}^2 > 10\sqrt{6} M_W^2 \frac{\sin^2\theta_W}{\cos\theta_W}
(\cos\Theta + 1) \sin\Theta \delta_3
\end{equation}
Since $\sin\Theta$ and $\delta_3$ are $\ll 1$\, we conclude that this bound
on $M_{Z'}$ is weaker than those obtained above from the rare kaon decays.


\section{Limits on the mixing angles between exotic and ordinary leptons.}

Here we use the formalism given in section 4 to analyze the mixing
between exotic and ordinary leptons. In particular, we estimate the
square of the mixing angles.

{}From the discussion of Section 2, and from the matrices $U_{CL(R)}$,
the submatrices $F_{L(R)}$ obtained in the model are
\be                                    \label{etreintaseis}
F_L = -\pmatrix{ \baselineskip=25pt
\frac{3w}{M^2_2}V''
&\frac w{M^2_2}\big(V'_1+V'_2+V'_3\big)
&\frac w{M^2_2}\big(V_1+V_2+V_3\big) \cr
\frac{3w}{M^2_2}V''
&\frac w{M^2_2}\big(V'_1+V'_2+V'_3\big)
&\frac w{M^2_2}\big(V_1+V_2+V_3\big) \cr
\frac{3w}{M^2_2}V''
&\frac w{M^2_2}\big(V'_1+V'_2+V'_3\big)
&\frac w{M^2_2}\big(V_1+V_2+V_3\big) \cr
}_L,
\ee
and $F_R=$
\be  \hspace{-2em}                       \label{etreintasiete}
\left( \footnotesize
\begin{array}{ccc}
\frac{\sqrt3w}{M_2}+ \frac{\sqrt3w}{M^3_2}\big(V_1V_3 +V'_1V'_3+V''^2\big)
&\frac{\sqrt3w}{M^3_2}\big(V_1V_2 +V'_1V'_2+V''^2\big)
&\frac{\sqrt3w}{M^3_2}\big(V_1^2 +V'_1{}^2_2+V''^2\big)    \\[1em]
 \frac{\sqrt3w}{M_2}+ \frac{\sqrt3w}{M^3_2}\big(V_1V_3 +V'_1V'_3+V''^2\big)
&\frac{\sqrt3w}{M^3_2}\big(V_1V_2 +V'_1V'_2+V''^2\big)
&\frac{\sqrt3w}{M^3_2}\big(V_1^2 +V'_1{}^2_2+V''^2\big)    \\[1em]
\frac{\sqrt3w}{M_2}+ \frac{\sqrt3w}{M^3_2}\big(V_1V_3 +V'_1V'_3+V''^2\big)
&\frac{\sqrt3w}{M^3_2}\big(V_1V_2 +V'_1V'_2+V''^2\big)
&\frac{\sqrt3w}{M^3_2}\big(V_1^2 +V'_1{}^2_2+V''^2\big)
\end{array}\right)_R.
 \ee
{}From these equations we can compute
$F^+_{L(R)}F_{L(R)}$.
The limits found for the mass of the horizontal gauge boson
which mediates FCNC, allow us to neglect the off diagonal terms, and
by the formalism of section 4, we can reparameterize the terms in the
diagonal; that is, we can write
\be
F^+_a F_a = \pmatrix{ \big(S^1_a\big)^2 &&0\cr
&\big(S^2_a\big)^2 \cr
0&&\big(S^3_a\big)^2\cr},
\qquad a=L,R,
\ee
where
   \def\ve#1{\frac{3w^2}{M^4_2}\big(#1_1+#1_2+#1_3\big)^2}
   \def\ss#1#2{\big(S^{#2}_{#1}\big)^2}
\be
\begin{array}{ccl}
     \ss L3 &= &\ve V, \\[.5em]
     \ss L2 &= &\ve {V'}, \\[.5em]
     \ss L1 &= &\frac{27w^2}{M^4_2} V''^2,
\end{array}
\ee
and
\be
     \big(S^1_R\big)^2 = \big(S^2_R\big)^2 = \big(S^3_R\big)^2 =
\begin{array}[t]{cl}
      &9 \l[ \frac w{M^2} +\frac w{M^3_2} (V_1V_3+V'_1V'_3+V''^2)\r]^2\\[.5em]
      &+9 \l[ \frac w{M^3_2} (V_1V_2  +V'_1V'_2 +V''^2)\r]^2\\[.5em]
      &+9 \l[ \frac w{M^3_2} (V_1^2  +V'_1{}^2  +V''^2)\r]^2.
\end{array}
\ee

With the constraint  $M_{Z'}\sim M_2$, and using
$V\sim m_e$, $V'\sim m_{\mu}$ and $V''\sim m_{\tau}$, we get
\def\y{\scriptstyle}
\[
\begin{array}{|c|c|c|c|c|}
\hline \y
M_{Z'} &\y \ss L1 &\y \ss L2 &\y \ss L3 &\y \ss Ri\\ \hline\y
19.0363\ \rm TeV &\y 654.443\times10^{-14} &\y  2.295\times10^{-14}
&\y 5.369\times
10)^{-19} &\y 24.836\times 10^{-5}  \\
\hline\y
34.5\ \rm TeV &\y 606.523\times10^{-14} &\y 2.127\times10^{-14} &\y
4.977\times10^{-
20} &\y 7.561\times10^{-5} \\ \hline
\end{array}
\]
where in the last column $i=1,2,3$.
\vskip.5cm

\section{Conclusions.}

We have analyzed some phenomenological implications
of the model\\
SU(6)$_L \otimes$ U(1)$_Y$ of unification of families
with the standard electroweak interactions. The model contains
exotic leptons and extra gauge bosons. The work has been focused on
FCNC and rare decays produced by three sources: the horizontal
interactions, the mixing between exotic and ordinary leptons, and
the mixing of the standard Z neutral gauge boson with a
horizontal gauge boson of SU(2)$_H$.

Given the complex structure of the mass matrices of fermions and
gauge bosons, we implement a mechanism to diagonalize these matrices
by using perturbation theory. To establish some limits on the scale
of the breaking of the
horizontal symmetry and limits for the mixing angles between exotic and
ordinary charged leptons, we have computed the rare decays
$K^{+}\longrightarrow\pi^{+}\mu^{+}e^{-}$ and
$K^{o}\longrightarrow\mu^{+}e^{-}$, as well as the muon decay
$\mu \longrightarrow e\bar{\nu}_{e}\nu_{\mu}$.

The results can be summarized as follows: a) The mass of the flavor
changing gauge boson is limited to values higher
than 34 Tev; b) the limits
on $S^2$, the square of the sine of the mixing angle between exotic and
ordinary charge leptons are stringent for the right sector
$(S^2_R \leq 7.5 \times {10}^{-5})$, and extremely stringent  for
the left sector $(S_L^2 \leq 6 \times {10}^{-12})$.

{\bf Acknowledgements}
This work was supported in part by CONACyT (Mexico).  A.Z. benefitted
from conversations with R. Shrock in the stimulating atmosphere of the
Aspen Center for Physics. We thank G. L\'opez- Castro for fruitful
conversations. R. G. acknowledges financial
support from the O.E.A. and W.A.P. from Colciencias (Colombia).

\pagebreak

\pagebreak

\noindent
Table 1

\noindent
Clasification of generators and gauge bosons of SU(6)$_L$. The numbers in
the first column denote (SU(2)$_L$, SU(2)$_H$) dimensionality.
\[
\begin{array}{|c|c|c|c|}
\hline
\rm Branching & \rm Class    & \rm Generators          &\rm  Gauge\ bosons \\
\hline
(3,1)     &\rm  standard & \sigma_1\otimes I_3 & W_1             \\
\cline{3-4}
          &          & \sigma_2\otimes I_3 & W_2             \\
\cline{3-4}
          &          & \sigma_3\otimes I_3 & W_3             \\ \hline
(1,3)     & \rm NUFDNC & I_2\otimes(\lambda_3+\sqrt{3}\lambda_8)/2 &
\Sigma_1 \\ \cline{2-4}
         & \rm FCNC    & I_2\otimes(\lambda_1+\lambda_6)/\sqrt{2}  &
Y_{H(1)} \\ \cline{3-4}
         &    &I_2\otimes(\lambda_2+\lambda_7)/\sqrt{2}   & Y_{H(2)} \\
\hline
(1,5)    & \rm NUFDNC  & I_2\otimes(\sqrt{3}\lambda_3-\lambda_8)/2 &
\Sigma_2 \\ \cline{2-4}
  & \rm FCNC & I_2\otimes(\lambda_1-\lambda_6)/\sqrt{2} & Y_{r(1)} \\
\cline{3-4}
 &   & I_2\otimes(\lambda_2-\lambda_7)/\sqrt{2} & Y_{r(2)} \\ \cline{3-4}
 &   & I_2\otimes\lambda_4 & Y_{3(1)}   \\ \cline{3-4}
 &   & I_2\otimes\lambda_5 & Y_{3(2)}   \\ \hline
(3,3) & \rm NUFDNC & \sigma_3\otimes(\lambda_3+\sqrt{3}\lambda_8)/2 & Z_3 \\
\cline{2-4}
& \rm FCNC & \sigma_3\otimes(\lambda_1+\lambda_6)/\sqrt{2} & E_{13(1)} \\
\cline{3-4}
&  &\sigma_3\otimes(\lambda_2+\lambda_7)/\sqrt{2} & E_{13(2)} \\ \cline{2-4}
 & \rm NUFDCC & \sigma_1\otimes(\lambda_3+\sqrt{3}\lambda_8)/2 & E_{12(1)} \\
\cline{3-4}
&   & \sigma_2\otimes(\lambda_3+\sqrt{3}\lambda_8)/2 & E_{12(2)} \\
\cline{2-4}
& \rm FCCC & \sigma_1\otimes(\lambda_1+\lambda_6)/\sqrt{2} & E_{14(1)} \\
\cline{3-4}
&  & \sigma_2\otimes(\lambda_1+\lambda_6)/\sqrt{2} & E_{14(2)} \\
\cline{3-4}
& & \sigma_1\otimes(\lambda_2+\lambda_7)/\sqrt{2} & F_{14(2)} \\
\cline{3-4}
& & \sigma_2\otimes(\lambda_2+\lambda_7)/\sqrt{2} & F_{14(1)} \\ \hline
(3,5) & \rm NUFDNC & \sigma_3\otimes(\sqrt{3}\lambda_3-\lambda_8)/2 &
Z_4 \\ \cline{2-4}
 & \rm FCNC & \sigma_3\otimes(\lambda_1-\lambda_6)/\sqrt{2} & F_{13(1)} \\
\cline{3-4}
& & \sigma_3\otimes(\lambda_2-\lambda_7)/\sqrt{2} & F_{13(2)} \\
\cline{3-4}
& & \sigma_3\otimes\lambda_4 & E_{15(1)} \\ \cline{3-4}
& & \sigma_3\otimes\lambda_5 & E_{15(2)} \\ \cline{2-4}
& \rm NUFDCC & \sigma_1\otimes(\sqrt{3}\lambda_3-\lambda_8)/2 & F_{12(1)} \\
\cline{3-4}
& & \sigma_2\otimes(\sqrt{3}\lambda_3-\lambda_8)/2 & F_{12(2)} \\ \cline{2-4}
& \rm FCCC & \sigma_1\otimes(\lambda_1-\lambda_6)/\sqrt{2} & G_{14(1)} \\
\cline{3-4}
& & \sigma_2\otimes(\lambda_1-\lambda_6)/\sqrt{2} & G_{14(2)} \\
\cline{3-4}
& & \sigma_1\otimes(\lambda_2-\lambda_7)/\sqrt{2} & H_{14(2)} \\ \cline{3-4}
& & \sigma_2\otimes(\lambda_2-\lambda_7)/\sqrt{2} & H_{14(1)} \\ \cline{3-4}
& & \sigma_1\otimes\lambda_4 & E_{16(1)} \\ \cline{3-4}
& & \sigma_2\otimes\lambda_4 & E_{16(2)} \\ \cline{3-4}
& & \sigma_1\otimes\lambda_5 & F_{16(2)} \\ \cline{3-4}
& & \sigma_2\otimes\lambda_5 & F_{16(1)} \\ \hline
\end{array}
\]
\pagebreak

{\bf Appendix}

{\bf Parameters $\theta_i$, $\beta_i$ and $\alpha_i$:}

\vspace{1cm}
$\alpha_1 = \left( \frac{V_1+V_2+V_3}{V^{\prime\prime}}
\right)/r^{\frac{3}{2}} $

\vspace{1cm}
$\alpha_2 = \left(
\frac{V_1^{\prime}+V_2^{\prime}+V_3^{\prime}}{V^{\prime\prime}}
\right)/r^{\frac{1}{2}} $

\vspace{1cm}
$\alpha_3 = \frac{1}{r^{\frac{1}{2}}}$

\vspace{1cm}
$\beta_1 = \left(
\frac{3(V_1+V_2+V_3)(V_1^{\prime}+V_2^{\prime}+V_3^{\prime})-
(V_1 V_1^{\prime}+V_2 V_2^{\prime}+V_3 V_3^{\prime})}
{({V_1^{\prime}}^2+{V_2^{\prime}}^2+{V_3^{\prime}}^2)-
3(V_1^{\prime}+V_2^{\prime}+V_3^{\prime})^2} \right)/r^{\frac{1}{2}}$

\vspace{1cm}
$\beta_2 = - \frac{1}{r^{\frac{1}{2}}}$

\vspace{1cm}
$\beta_3=\alpha_2$

\vspace{1cm}
$\theta_1 = 1$

\vspace{1cm}
$\theta_2 = \frac{\beta_1}{r^{\frac{1}{2}}}$

\vspace{1cm}
$\theta_3 = - \left( \frac{\alpha_1}{r^{\frac{1}{2}}} \right) -
\beta_1 \left( \frac{V_1^{\prime}+V_2^{\prime}+V_3^{\prime}}
{V^{\prime\prime} r^{\frac{1}{2}}} \right)$

\vspace{5mm}
where

\begin{center}
$r=1 +
\frac{(V_1^{\prime}+V_2^{\prime}+V_3^{\prime})^2}{{V^{\prime\prime}}^2}$
\end{center}
\pagebreak

\vspace*{2.5in}

Fig.~1 Tree level diagram for $K^+\longrightarrow \pi^+e^-\mu^+$ through a
horizontal gauge boson $Z'$.
\vfill

Fig.~2  Tree level diagram for $K^0_L\longrightarrow e^-\mu^+$ through a
horizontal gauge boson $Z'$.
\vskip.5cm

\vspace*{2.5in}

Fig.~3 Diagram which contribute to the $\mu\longrightarrow e \bar\nu_e
\nu_\mu$ in SU(6)$_L\otimes$U(1)$_Y$ through the weak gauge boson $W$.

Fig.~4 Diagram which contribute to the $\mu\longrightarrow e \bar\nu_e
\nu_\mu$ in SU(6)$_L\otimes$U(1)$_Y$ through the horizontal  gauge boson $Z'$.

\end{document}